\documentclass{appolb}%
\usepackage{epsfig}
\usepackage{amsmath}
\usepackage{amsfonts}
\usepackage{amssymb}
\usepackage{graphicx}%

\begin{document}

\title{Saturation and Scaling of Multiplicity, Mean $p_{\text{T}}$ and $p_{\text{T}}$
Distributions from $200~\mathrm{GeV}\leq\sqrt{s}\leq7~\mathrm{TeV} $.}
\author{Larry McLerran \address{BNL and Riken Brookhaven Center, Upton, NY }
\and Michal Praszalowicz \address{
M. Smoluchowski Institute of Physics, Jagellonian University, Reymonta 4,
30-059 Krakow, Poland} }
\maketitle

\begin{abstract}
The multiplicity, average transverse momentum, and charged particle transverse
momentum distributions have recently been measured in LHC experiments. The
multiplicity and average transverse momentum grow with beam energy. Such
growth is expected in the theory of the Color Glass Condensate, a theory that
incorporates the physics of saturation into the evolution of the gluon
distribution. We show that the energy dependence of the $p\overline{p}$ data
and the LHC data for $pp$ scattering at $\sqrt{s}\geq200$ GeV may be simply
described using a minimal amount of model input. Such a description uses
parameters consistent with the Color Glass Condensate descriptions of HERA and
RHIC experimental data.

\end{abstract}


\section{Introduction}

The first LHC data has been released on total charged particle multiplicity as
a function of energy, average transverse momenta of charged particles as a
function of energy \cite{:2009dt}-\cite{Khachatryan:2010us} and transverse
momentum of charged particles as a function of charged particle multiplicity
\cite{Aad:2010rd}. The generic features of the experimental data are that the
charged particle pseudo-rapidity densities rise as a power of energy,
$dN/d\eta\sim E^{0.23}$, and that the average transverse momentum rises with
both energy and charged particle multiplicity. This behavior has a natural
explanation within the theory of saturation and the Color Glass Condensate
\cite{Gribov:1984tu} -\cite{Ayala:1995hx}. Within this theory, the total
multiplicity of produced particles is computable \cite{Kovner:1995ts}%
-\cite{Kovner:1995ja}. Assuming local parton hadron duality
\cite{Dokshitzer:1991eq}, the initially produced gluon multiplicity is
proportional to the charged particle multiplicity. The pseudo-rapidity
multiplicity density can be expressed in terms of the saturation momentum as
\cite{Mueller:1999fp}-\cite{Kharzeev:2004if}:
\begin{equation}
{\frac{1}{\sigma}}{\frac{{dN}_{\text{ch}}}{{d\eta}}}= {\frac{{\rm const.}}{{\alpha
_{S}(Q_{\mathrm{sat}})}}}Q_{\mathrm{sat}}^{2}%
\end{equation}
The strong coupling constant is evaluated at the saturation momentum scale.

The saturation momentum scale is proportional to the transverse gluon density.
Evolution equations that build in the effects of high gluon density have been
derived in Refs. \cite{Balitsky:1995ub}-\cite{Weigert:2000gi}. The dependence
of the saturation momentum on fractional gluon momentum may be determined from
such considerations \cite{Triantafyllopoulos:2002nz}. It has been shown that
the generic features of such a description of saturation can describe the HERA
data on inclusive and diffractive deep inelastic
scattering\cite{GolecBiernat:1998js}-\cite{Albacete:2009ps}, and correctly
predicts observed scaling properties of experimental data\cite{Stasto:2000er}%
-\cite{Iancu:2002tr}. The results of this analysis relevant for our purposes
is that the saturation momentum at $x$ values appropriate for the LHC scales
with $x$ as $Q_{\mathrm{sat}}^{2}\sim1/x^{\lambda}$ where $\lambda\sim0.2-0.3$

There have been detailed saturation based predictions and descriptions of the
recent results from the LHC \cite{Kharzeev:2004if}, \cite{Rezaeian:2009it}-\cite{Levin:2010dw}. The
goal of this paper is not to improve upon the descriptions provided in these
works. Our goal is to show that the simplest generic features of saturation
based descriptions are adequate to quantitatively describe data on the
dependence particle multiplicities on energy and the dependence of such
average transverse momentum upon multiplicity and beam energy. We will also
argue that there is an approximate geometric scaling of transverse momenta
distributions measured at LHC energies.

\section{
Color Glass Condensate Description of the LHC data}

To reduce the Color Glass Condensate description to its simplest possible
form, we will assume that the density of produced charged particles per unit
pseudo-rapidity scales as
\begin{equation}
{\frac{{dN}_{\text{ch}}}{{d\eta}}}=\kappa Q_{\mathrm{sat}}^{2}=AE^{\lambda
}\label{eq:NchvsE}%
\end{equation}
Here the energy is measured in units of TeV.

\begin{figure}[ptb]
\centering
\includegraphics[scale=0.90]{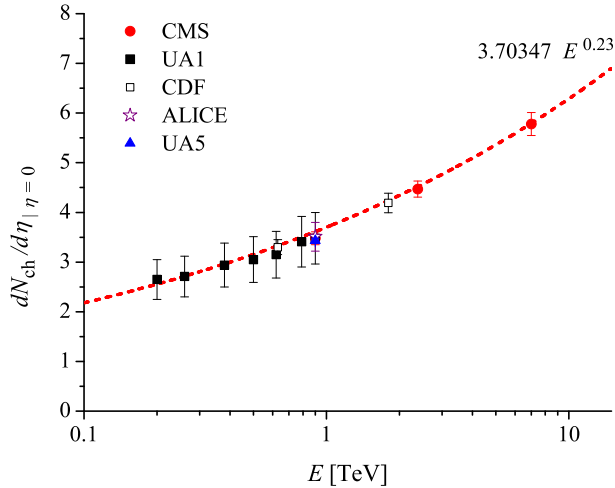}\caption{Charged particle rapidity
density as a function of energy compared to a power law $3.7043~E^{0.23}$ Data
from the LHC\cite{:2009dt}-\cite{Khachatryan:2010us},
UA(1)\cite{Albajar:1989an}, UA(5)\cite{Alner:1986xu} and CDF \cite{Abe:1989td}%
-\cite{Abe:1988yu}. }%
\label{Nch}%
\end{figure}The parameters $\lambda$ and $A$ can be determined by a fit to the
LHC data. In Fig.~\ref{Nch}, an excellent fit is shown to the data that also
includes lower energy data for proton-anti-proton scattering. (This agreement
with the $p\overline{p}$ data is a little surprising since there should be
some small difference between $pp$ and $p\overline{p}$ scattering at the
energies of interest.)

\begin{figure}[ptb]
\centering
\includegraphics[scale=0.90]{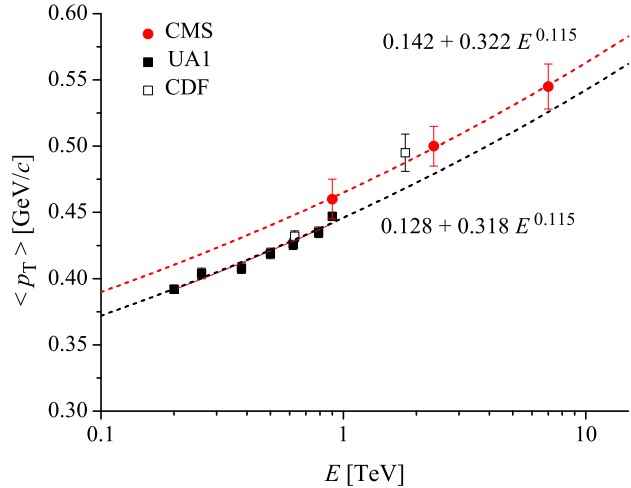}\caption{Average transverse momentum as a
function of energy compared to a power law $B+CE^{0.115}$ Data from the
LHC\cite{:2009dt}-\cite{Khachatryan:2010us}, UA(1)\cite{Albajar:1989an},
UA(5)\cite{Alner:1986xu} and CDF \cite{Abe:1989td}-\cite{Abe:1988yu}. }%
\label{ptvse}%
\end{figure}

On dimensional grounds, at very high energy, we expect the average transverse
momentum will be proportional to the saturation momentum. Since this term does
not entirely dominate the contribution to the transverse momentum at
accessible energies, we add a constant, so that the result has a reasonable
low limit at lower energy. In this case, there is a small difference seen in
the data between $pp$ and $p\overline{p}$ scattering. The power law growth
should be, however, universal. We fit the average $p_{\mathrm{T}}$ with the
following form
\begin{equation}
\left\langle p_{\mathrm{T}}\right\rangle =B+CE^{\lambda/2}.\label{ptElam}%
\end{equation}
The results of such a fit are shown in Fig.~\ref{ptvse}. We see that indeed
the power of the energy and its coefficient (within errors which we do not
quote here) are identical both for $pp$ and $p\overline{p}$ data. Similar
form of the average $\left\langle p_{\mathrm{T}}\right\rangle$ has been
recently postulated in Ref. \cite{Troshin:2010yy} with
higher power of $E$ equal to 0.414. However, the recent $7$~TeV CMS
point is far below their curve.

If the saturation momentum is the only scale that controls $p_{\mathrm{T}}$
distributions, on dimensional grounds, these distributions should have a
geometrical scaling
\begin{equation}
{\frac{1}{\sigma}}{\frac{{dN}_{\text{ch}}}{{d\eta d^{2}p_{\mathrm{T}}}}%
}=F\left(  {\frac{{p_{\mathrm{T}}}}{{Q_{\mathrm{sat}}(p_{\mathrm{T}}/\sqrt
{s})}}}\right)  .
\end{equation}
This means that there is a universal function of the scaling variable%
\begin{equation}
\tau={\frac{{p_{\mathrm{T}}^{2}}}{{Q_{\mathrm{sat}}^{2}(p_{\mathrm{T}}%
/\sqrt{s})}}=}\frac{{p_{\mathrm{T}}^{2}}}{1\,\text{GeV}^{2}}\left(
\frac{{p_{\mathrm{T}}}}{\sqrt{s}\times10^{-3}}\right)  ^{\lambda
}\label{eq:tau}%
\end{equation}
that describes the data at different energies (where $p_{\text{T}}$ and
$\sqrt{s}$ are in GeV). This is of course limited in the range of
$p_{\text{T}}$, so that one is not probing quark and gluon distributions
outside the saturation region. By a rescaling of variables, one can check if
the data from CMS falls on a universal scaling curve. Since the data points of
the CMS $p_{\text{T}}$ distributions are not yet publicly accessible we shall
use throughout this paper an analytical parametrization in terms of Tsallis fit
\cite{Tsallis:1987eu} as given in Refs.~\cite{Khachatryan:2010xs,Khachatryan:2010us}:%
\begin{equation}
\frac{{dN}_{\text{ch}}}{{d\eta d^{2}p_{\mathrm{T}}}}=C\frac{p}{E}\frac
{{dN}_{\text{ch}}}{{d\eta}}\left(  1+\frac{E_{T}}{nT}\right)  ^{-n}%
\label{Tsalis}%
\end{equation}
where $E_{\mathrm{T}}=\sqrt{m_{\pi}^{2}+p_{\text{T}}^{2}}-m_{\pi}$ and%
\begin{equation}%
\begin{array}
[c]{c|cc}%
\sqrt{s}\,\text{[TeV]} & T\,\text{[GeV]} & n\\\hline
0.9 & 0.130 & 7.7\\
2.36 & 0.140 & 6.7\\
7.0 & 0.145 & 6.6
\end{array}
\end{equation}

\begin{figure}[tbh]
\centering
\includegraphics[scale=0.65]{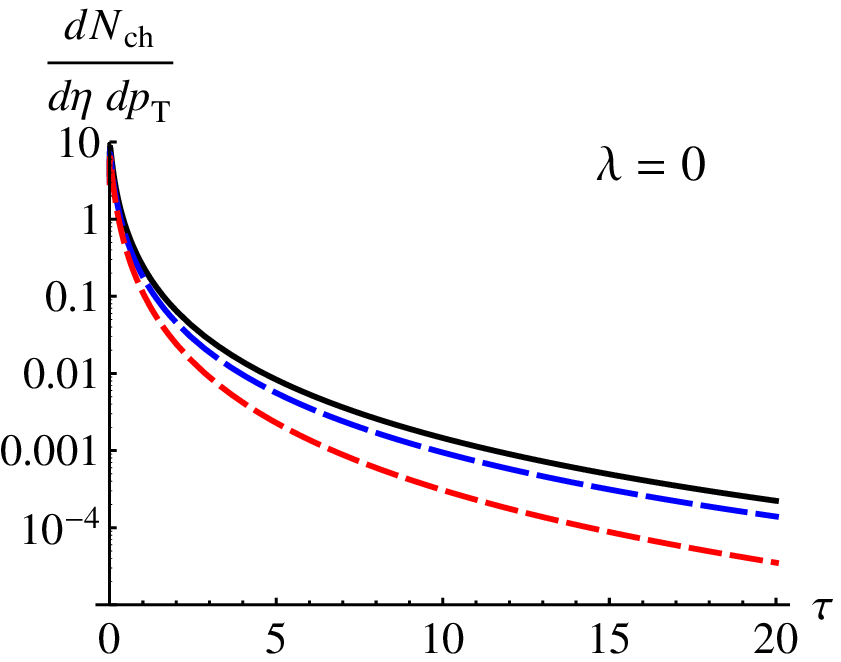} \quad
\includegraphics[scale=0.65]{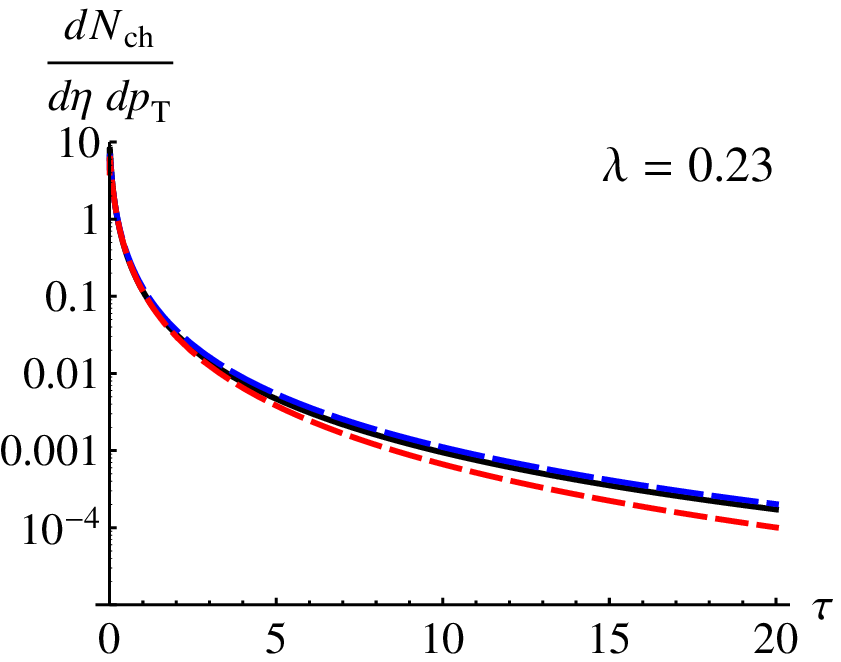}\caption{ The CMS data for
transverse momentum distributions on the left as functions of $p_{\mathrm{T}%
}^{2}$ (note that for $\lambda=0$ scaling variable $\tau=p_{\mathrm{T}}%
^{2}/(1~\mathrm{GeV}^{2}))$. On the right, the same $p_{\mathrm{T}}$
distribution rescaled in terms of the scaling variable $\tau=p_{\mathrm{T}%
}^{2}/Q_{\mathrm{sat}}^{2}(p_{\mathrm{T}}/\sqrt{s})$. }%
\label{dNch-dtau}%
\end{figure}

Up to about $4-6~$GeV, the limit of the available data, such a scaling
relation is satisfied, as shown in Fig.~\ref{dNch-dtau}. 

\begin{figure}[tbh]
\centering
\includegraphics[scale=0.65]{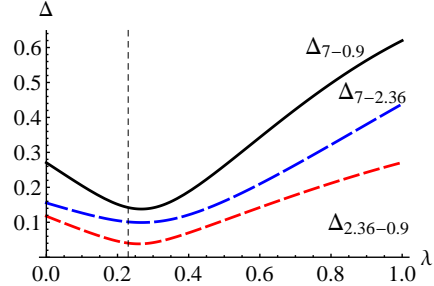} \quad\caption{Normalized square
deviations between scaled $p_{\mathrm{T}}$ distributions for different CMS
energies as function of saturation parameter $\lambda$. Optimal DIS
$\lambda=0.23$ is marked by a thin vertical line. }%
\label{Deltas}%
\end{figure}

The scaling plot in Fig.~\ref{dNch-dtau} has been obtained by using power
$\lambda$ in Eq.(\ref{eq:tau}) fixed from the DIS data at HERA. It is
interesting to see whether this is also the optimal power for hadron-hadron
scattering. To this end we compute the mean deviation of the scaled
$p_{\text{T}}$ distributions for different energies%
\begin{equation}
\sigma_{E_{1}-E_{2}}^{2}=%
{\displaystyle\int\limits_{0}^{\tau_{\text{max}}}}
\left(  \left.  \frac{{dN}_{\text{ch}}}{{d\eta d\tau}}\right\vert _{E_{1}%
}-\left.  \frac{{dN}_{\text{ch}}}{{d\eta d\tau}}\right\vert _{E_{2}}\right)
^{2}d\tau
\end{equation}
and normalizing them to the sum%
\begin{equation}
s_{E_{1}-E_{2}}=%
{\displaystyle\int\limits_{0}^{\tau_{\text{max}}}}
\left(  \left.  \frac{{dN}_{\text{ch}}}{{d\eta d\tau}}\right\vert _{E_{1}%
}+\left.  \frac{{dN}_{\text{ch}}}{{d\eta d\tau}}\right\vert _{E_{2}}\right)
d\tau
\end{equation}
we define quantities%
\begin{equation}
\Delta_{E_{1}-E_{2}}=\frac{\sigma_{E_{1}-E_{2}}}{s_{E_{1}-E_{2}}%
}\label{eq"Delta}%
\end{equation}
that are plotted in Fig.~\ref{Deltas}. We see that the minima obtained with
the Tsallis fit (\ref{Tsalis}) are rather shallow and include the optimal value
of $\lambda$ obtained from DIS, although the preferred value would be slightly
bigger. We checked, however, that the higher value of $\lambda$ is
incompatible with the energy dependence of charged multiplicity shown in
Fig.~\ref{Nch}. We find this agreement (note that we use Tsallis
parametrization instead of real data) as a strong support of the applicability
of geometric scaling to hadron-hadron scattering. 

\begin{figure}[tbh]
\centering
\includegraphics[scale=0.90]{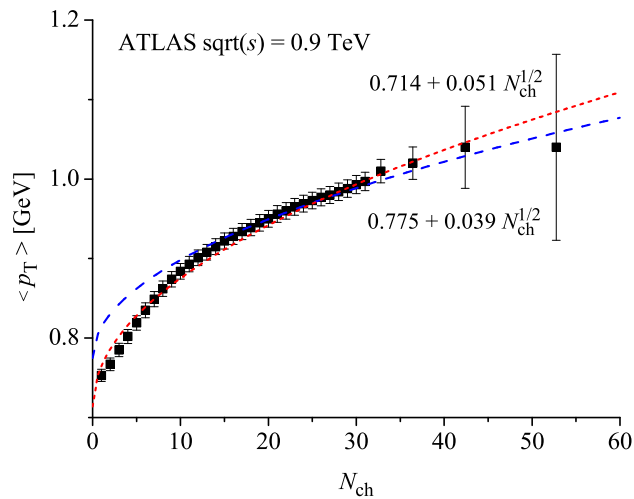} \quad\caption{ATLAS data compared to
the square root fit of Eq.(\ref{sqfit}). Short-dashed red curve corresponds to
the fit over the whole range of multiplicities, whereas long-dashed blue fit
is restricted to $N_{\mathrm{ch}}>15$.}%
\label{ptvsNch}%
\end{figure}

In the ATLAS experiment, the average $p_{\mathrm{T}}$ of events with various
multiplicities was computed. There was a transverse momentum cutoff of
$p_{\mathrm{T}}\geq500$~ MeV. We expect as in Eq.~(\ref{ptElam}), that the
average transverse momentum will be
\begin{equation}
\left\langle p_{\mathrm{T}}\right\rangle =C+D\sqrt{N_{\mathrm{ch}}%
}\label{sqfit}%
\end{equation}
Since the CMS and ATLAS cuts are different we simply show
in Fig.~\ref{ptvsNch} that such functional form
provides a good description of the experimental data. We have presented
two fits: one to the whole region of available multiplicities
and the second one for $N_{\rm ch}>15$.
The latter choice is dictated by the slight change of the curvature of the
data around $N_{\rm ch}\sim 10$.

\section{Predictions for Higher Energy}

\begin{figure}[ptb]
\centering
\includegraphics[scale=0.90]{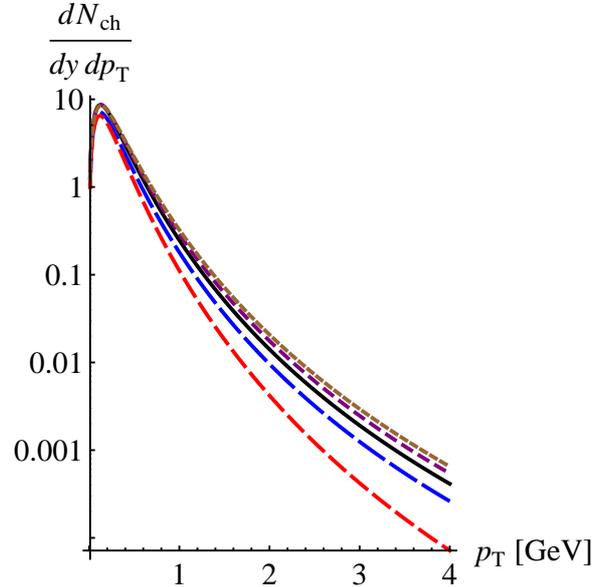}
\caption{CMS  particle spectra in terms of Tsallis
parametrization of Refs.~\cite{Khachatryan:2010xs,Khachatryan:2010us}
for different energies. Solid line and long and short dashed lines below
correspond to $\sqrt{s}=7,~2.36$ and $0.9$ TeV respectively. Dashed
and short dashed lines above the solid line correspond to our predictions
for $\sqrt{s}=10$ and $14$~TeV respectively.
 }%
\label{fig:spectra}%
\end{figure}

Using the scaling analysis in this paper we can make predictions for the
multiplicity per unit rapidity, average transverse momentum and transverse
momentum distributions at higher LHC energies. Our predictions for the
multiplicity per unit rapidity are shown in fact in Fig.~\ref{Nch}. In order
to estimate roughly the error of that fit we simply propagated the
experimental error of the 7 TeV point with the help of Eq.(\ref{eq:NchvsE})
obtaining $\left.  dN_{\mathrm{ch}}/d\eta\right\vert _{\eta=0}=6.29\pm0.25$
and $6.80\pm0.27$ for $\sqrt{s}=10$ and $14$~TeV respectively. In a similar
way we have estimated average $\left\langle p_{\mathrm{T}}\right\rangle
=0.562\pm0.017$ and $0.579\pm0.017$. By minimizing $\Delta_{10-7}$ and
$\Delta_{14-7}$ with respect to parameters $T$ and $n$ of the Tsallis formula
(\ref{Tsalis}) we have obtained $p_{\mathrm{T}}$ distributions at $\sqrt
{s}=10$ and $14~$TeV that are shown in Fig.~\ref{fig:spectra}. The
corresponding Tsallis parameters read: $T_{10}=0.153$, $n_{10}=6.6$ for
$\sqrt{s}=10$~TeV and $T_{14}=0.162$, $n_{14}=6.7$ for $14$~TeV.

The scaling behavior we see in $pp$ collisions can be used to estimate initial
state effects for the heavy ion collisions. Such effects might be very
important for measuring jet quenching effects, once the $A$ dependence of the
saturation momentum is established at LHC energy. Note that the asymptotic
behavior of the Tsallis fit for high energies is controlled by the parameter
$n^{2} T/p_{\rm T}$. If $T$ scales as $A^{1/3}$, then the asymptotic limit is
obtained only at very high transverse momentum values, suggesting that
saturation effects can influence transverse momentum distributions out to very
high large values. This may influence experimental studies of jet quenching as
an attempt to extract properties of the Quark Gluon Plasma.

\section*{Acknowledgments}

We wish to thank the organizers of the Krakow School of Theoretical Physics in
Zakopane, Poland where this work was initiated. We particularly thank Andrzej
Bialas and Krzysztof Golec Biernat for insightful comments, and Dave Charlton
for a very clear presentation of the results from the LHC experiments. We also
thank Eugene Levin and Raju Venugopalan for their critical reading of the manuscript. The
research of L. McLerran is supported under DOE Contract No. DE-AC02-98CH10886.

\end{document}